\pdfoutput=1
\documentclass[aps,twocolumn,showpacs,superscriptaddress,prl]{revtex4-1}

\usepackage[pdftex]{graphicx}
\usepackage[pdftex]{epsfig}
\usepackage{epstopdf}
\usepackage{color}
\usepackage{amssymb,amsmath}
\usepackage[hypertexnames=false]{hyperref}
\usepackage{algorithm}
\usepackage[noend]{algpseudocode}
\usepackage{braket}

\def\figref#1{Fig.~\ref{#1}}
\def\figsref#1#2{Figs.~\ref{#1} and\ref{#2}}
\def\eqnref#1{Eq.~\ref{#1}}
\def\eqnsref#1#2{Eqs.~\ref{#1} and~\ref{#2}}
\DeclareMathOperator*\lsp{span}
\DeclareMathOperator\arctantwo{atan2}

\def\estphase{\theta}           % the phase being estimated
\def\xyphase{\varphi}           % the azimuthal angle in the xy plane
\def\glophase{\chi}             % the unobserved global phase
\def\estunitary{\mathcal{W}}    % the unitary whose phases we're estimating
\def\reps{k_{g}}
\def\errc{\mathcal{E}_\text{c}}
\def\errl{\mathcal{E}_\text{l}}
\def\errp{\mathcal{E}_\text{p}}
\def\Bselect{\mathcal{B}}

\begin{document}

\title{Evaluating energy differences on a quantum computer with robust phase estimation}
\author{A.E. Russo}
\author{K.M. Rudinger}
\affiliation{Center for Computing Research, Sandia National Laboratories, Albuquerque NM 87185}
\author{B.C.A. Morrison}
\author{A.D. Baczewski}
\affiliation{Center for Computing Research, Sandia National Laboratories, Albuquerque NM 87185}
\affiliation{Center for Quantum Information and Control (CQuIC), University of New Mexico, Albuquerque NM, USA}

\begin{abstract}
We adapt the robust phase estimation algorithm to the evaluation of energy differences between two eigenstates using a quantum computer.
This approach does not require controlled unitaries between auxiliary and system registers or even a single auxiliary qubit.
As a proof of concept, we calculate the energies of the ground state and low-lying electronic excitations of a hydrogen molecule in a minimal basis on a cloud quantum computer.
The denominative robustness of our approach is then quantified in terms of a high tolerance to coherent errors in the state preparation and measurement.
Conceptually, we note that all quantum phase estimation algorithms ultimately evaluate eigenvalue differences.
\end{abstract}

\maketitle

\textit{Introduction.---}
The assessment of energy differences, rather than total energies, is ubiquitous throughout physics.
Assessing whether there is a gap between the ground and first excited state of a particular Hamiltonian is related to outstanding problems in condensed matter~\cite{haldane1983nonlinear} and high energy physics~\cite{jaffe2006quantum}, and it is even at the heart of deep connections between many-body physics and theoretical computer science~\cite{cubitt2015undecidability}.
More generically, myriad spectroscopic techniques ultimately compare the energies of two or more eigenstates of a single Hamiltonian as one among many identifying features of a particular piece of matter.
This paper is concerned with using a quantum computer for this purpose.
We indicate the Hamiltonian of interest as $H$ with $N=2^n=\dim{H}$.
The state $|0\rangle$ refers to a computational basis state for an $n$-qubit register unless otherwise indicated.

\begin{figure*}[ht]
\includegraphics{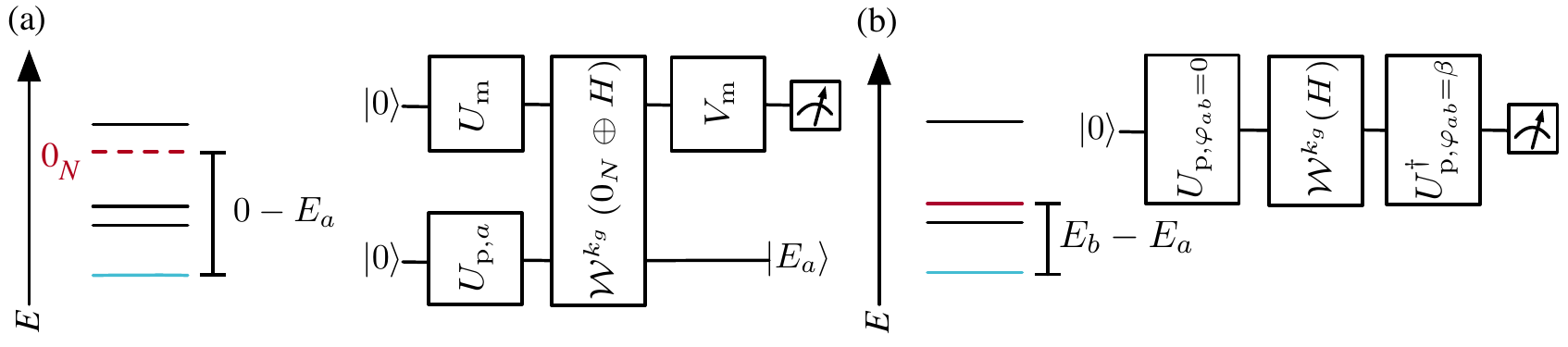}
 \caption{A comparison of prototypical QPE and RPE circuits.
 (a) In QPE with one auxiliary qubit, the system register is first prepared in the $a$th eigenstate of $H$ ($U_{\text{p},a}$) while a change of basis is applied to the auxiliary qubit ($U_m$).
 $\reps$ applications of $\estunitary$ are applied using the auxiliary qubit as a control ($\estunitary^{\reps}(0_N\oplus H)$).
 The basis of the auxiliary qubit is changed ($V_m$) into that in which it is measured, directly extracting a single bit of $E_a$.
 The unitary whose phase is being estimated acts on a $2N$-dimensional Hilbert space in which the auxiliary qubit provides an $N$-fold degenerate zero-energy subspace relative to which a phase difference maps onto any of the $N$ energy eigenvalues of $H$.
 (b) In RPE there is no auxiliary qubit.
 First, a uniform superposition of the $a$th and $b$th eigenstates of $H$ is prepared ($U_{\text{p},\xyphase_{ab}=0}$).
 Then $\reps$ applications of $\estunitary$ are applied ($\estunitary^{\reps}(H)$).
 The uniform superposition is unprepared ($U^\dagger_{\text{p},\xyphase_{ab}=\beta}$) and all qubits are measured, resulting in a sample from $P_{\text{c}}$ for $\beta=0$ and a sample from $P_{\text{s}}$ for $\beta=\pi/2$.
 Here, $P_{\text{c}}$ and $P_{\text{s}}$ are probability distributions that encode the energy difference of interest, defined in Eq.~\ref{eq:distributions}.
 The unitary whose phase is being estimated acts on an $N$-dimensional Hilbert space so that only energy differences between eigenstates of $H$ can be extracted from phases.
 \label{fig:summary}}
\end{figure*}

By repeatedly preparing particular superpositions of two energy eigenstates, allowing them to undergo a unitary evolution $\estunitary(H)$~\cite{lloyd1996universal,childs2010relationship,berry2015simulating,babbush2018encoding}, undoing the preparation, and measuring in the computational basis (see \figref{fig:summary}b), we can infer the difference in energy between the two eigenstates without the need for auxiliary qubits~\footnote{Throughout we use the phrase ``auxiliary qubit'' in place of the phrase ``ancilla qubit'' following the etymological concerns raised in~\cite{wiesner2017careless} and an alternative proposed in~\cite{puzzuoli2018entanglement}.} or controlled unitary operations.
This differs from other approaches to quantum phase estimation (QPE)~\cite{kitaev1995quantum} that use one or more auxiliary qubits to provide a ground reference for the phase accumulated on the register encoding the physical system~\cite{griffiths1996semiclassical, knill2007optimal, wiebe2016efficient, svore2018faster, low2019q, o2019quantum}.
This procedure is inspired by the robust phase estimation (RPE) algorithm that was introduced for the purposes of characterizing and calibrating the phase (i.e., rotation angle) of a single-qubit gate~\cite{kimmel2015robust}.

A common form for $\estunitary(H)$ is some approximation to the exponential map that describes Hamiltonian evolution for a fixed time, though it might take other forms for which the associated phase is a known function of the eigenvalues.
The physical significance of $\estunitary(H)$ is a consequence of encoding the degrees of freedom of some system of interest into the Hilbert space of $n$ qubits.
While we consider the specific encoding of interacting electrons in a molecular system~\cite{aspuru2005simulated}, we note that our results can be extended to others including those relevant to nuclear matter~\cite{dumitrescu2018cloud}, quantum field theories~\cite{jordan2012quantum}, and spin systems~\cite{childs2018toward}.
We label the ground state of $H$ with its eigenvalue, $\ket{E_0}$, and indicate the $a$th eigenstate above it as $\ket{E_a}$.

In fact all forms of phase estimation, with or without auxiliary qubits, are not simply eigenvalue estimation but eigenvalue \textit{difference} estimation.
The operations $\estunitary(H)$ and $\estunitary(H+\alpha I)$ are identical up to an undetectable global phase, $\exp(i\glophase(\alpha))$, where the form of $\glophase$ depends on $\estunitary$~\footnote{For example, in approaches to simulation based on Trotterization $\estunitary(H)\approx\exp(iH t)$ and thus $\glophase(\alpha)=\alpha t$}.
In order to actually estimate the phase of an eigenstate of $\estunitary$, one must have access to a known \emph{reference} energy level.
$\Lambda(\estunitary)$, a singly-controlled version of $\estunitary$, is generated by a Hamiltonian of the form $0_N\oplus H$, where $0_N$ is the $N\times N$ zero matrix.
The $N$-fold degenerate zero-energy subspace created by $0_N$ allows for the estimation of the phase of \textit{any} of the eigenstates of $H$ relative to these reference eigenstates (see \figref{fig:summary}a).
This is the structure of \emph{most} QPE implementations, which we henceforth generically refer to as QPE algorithms with auxiliary qubits.
Part of what distinguishes RPE is that instead of relying on the auxiliary register to relativize the phase of the Hamiltonian evolution, the relative phase is accumulated between two energy eigenstates in a uniform superposition.
This allows us to avoid the use of an auxiliary register and controlled unitaries at the cost of requiring a more complicated state preparation.

\begin{figure}[t]
\includegraphics[width=\columnwidth]{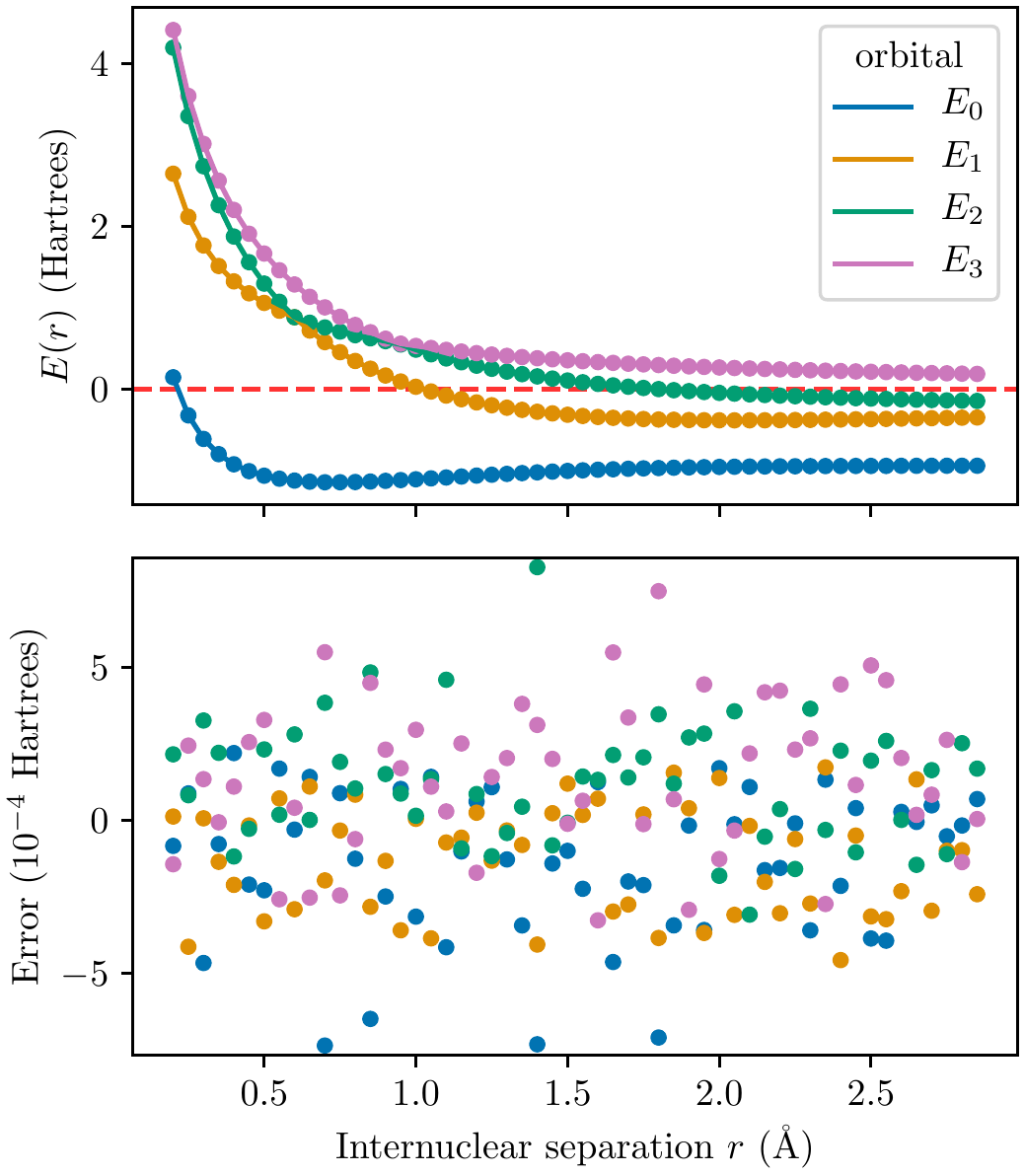}
 \caption{Verification of RPE for evaluating energy differences in a molecule using a cloud quantum computer.
 (Top) The first four energy levels of H$_2$ in a minimal basis, as evaluated using RPE on IBM Vigo (dots) and diagonalization on a classical computer (lines).
 (Bottom) The error in the first four energy levels relative to the result evaluated on a classical computer.
 \label{fig:h2_shibboleth}}
\end{figure}

Three strengths of approaches to QPE with auxiliary qubits are (i) the relativization of the phase accumulated on the 1 branch of the auxiliary register to the 0 branch, (ii) the projection of the system register onto an energy eigenstate after a single round, and (iii) the ability to continually reuse that state in subsequent rounds without having to prepare it again.
Point (i) is a critical advantage if one needs to know absolute energies, but not actually essential if one is strictly interested in measuring energy differences.
Further, if we have access to the trace of the Hamiltonian over a $M$-dimensional subspace it is still possible to reconstruct the absolute energies if $M-1$ independent pairwise energy differences are measured within that subspace.
This is evident in the experimental results in \figref{fig:h2_shibboleth}.
Points (ii) and (iii) are critical advantages if the depth of the state preparation unitary exceeds that of the longest Hamiltonian evolution unitary, noting that the depth of the Hamiltonian evolution unitaries for RPE will be reduced by merit of their not needing to be controlled unitaries.

Not only does RPE offer overall circuit depth improvement, but it offers significant savings in the total number of CNOTs/entangling operations required, which are a primary bottleneck in current hardware, given their relatively low fidelities (compared to single-qubit gates) \cite{bruzewicz2019trapped,kjaergaard2019superconducting}.
Given access to a gate-level description of a circuit $\mathcal{S}$ that implements $\mathcal{W}(H)$ (using only arbitrary local gates and CNOTs), the most straightforward way to implement $\mathcal{W}(0_N \oplus H)$ is to simply turn every gate $\mathcal{G}$ in $\mathcal{S}$ into its singly-controlled version $\Lambda(\mathcal{G})$.
Though clever compilation schemes \cite{barenco1995elementary, khatri2019quantum, venturelli2019quantum, maslov2017basic, botea2018complexity, booth2018comparing} may offer non-trivial improvements, if $\mathcal{G}$ contains $s$ single-qubit gates and $t$ CNOTs, it can be shown \cite{song2002optimal,shende2008cnot} the overall CNOT cost of implementing $\Lambda(\mathcal{G})$ may be as bad as $6t+2s$.
While this difference may be trivial on error-corrected hardware, it can be non-trivial for uncorrected qubits.

One might ask whether the need to repeat the potentially erroneous state preparation and measurement (SPAM) due to a lack of projection onto an energy eigenstate after a single round of auxiliary-qubit-free phase estimation is a limiting factor.
A central result of this paper is the observation that the robustness of RPE manifests as a surprisingly high tolerance to errors in the requisite SPAM.
This suggests conditions for which this approach to phase estimation might be advantageously employed for quantum simulation.
In particular, an advantage might be realized in the intermediate-term where adiabatic~\cite{farhi2001quantum, aspuru2005simulated} or filtering-based~\cite{poulin2009preparing, ge2019faster, lin2020near} state preparation can be replaced by precompiled state preparation circuits that exploit classical tractability and do not appreciably contribute to the total circuit depth (see Supplemental Materials).

\textit{Methods.---}
In essence, RPE may be thought of as a combination of Ramsey and Rabi oscillation experiments with logarithmic spacing in the number of gate repetitions~\footnote{By ``logarithmic spacing'' we mean that the circuit depths are, e.g., $1, 2, 4, 8\ldots$, as opposed to, e.g., $1, 2, 3, 4\ldots$, i.e., the circuit depths are spaced uniformly on a logarithmic scale.}.
This allows the phase of the gate to be learned with Heisenberg-like scaling in accuracy, without requiring any entanglement or auxiliary qubits.
Additionally, RPE will still produce accurate phase estimates even when there is a significant amount of error in any of the constituent circuits' state preparations, measurements, or gates.
Accordingly, RPE has been demonstrated in experimental systems to yield highly accurate phase estimates \cite{rudinger2017experimental} while being robust against various noise channels \cite{meier2019testing}.

While RPE concerns itself with estimating a single-qubit gate's phase, (e.g., the angle $\theta$ in the gate $R_x(\theta)=\exp(-i\theta\sigma_x/2)$), this phase is actually the difference between the two eigenvalues of the Hamiltonian which generates the unitary rotation \footnote{For the example of $R_x(\theta)=\exp(-i\theta\sigma_x/2)$, the Hamiltonian which generates $R_x$ is $\tfrac{\theta\sigma_x}{2}$; by inspection the Hamiltonian eigenvalue difference is $\theta$.}.
This principle can be generalized to unitary maps of dimension greater than two, allowing for the difference between two eigenvalues of an arbitrary Hamiltonian to be estimated using RPE.

To adapt RPE to higher dimensions one simply needs implementations of (i) $\estunitary(H)$ and (ii) a state preparation unitary,
\begin{equation}
    U_{\text{p},\xyphase_{ab}=\beta}|0\rangle = \frac{1}{\sqrt{2}}\left(\ket{E_a}+e^{i\beta}\ket{E_b}\right) = \ket{\xyphase_{ab}=\beta},
\end{equation}
where we will specifically need $U_{\text{p},\xyphase_{ab}}$ for two values of $\xyphase_{ab}$ that are separated by $\pi/2$ radians.
The energy difference between eigenstates $a$ and $b$ is related to a relative phase, $\estphase_{ab}$ mod $2\pi$, accumulated while evolving with $\estunitary(H)$ for a particular time interval that is absorbed into the units.
This relative phase is encoded in the probability distributions
\begin{subequations}
\begin{align}
    P_\text{c}(\reps\estphase_{ab}) &= |\langle 0| U^{\dagger}_{\text{p},\xyphase_{ab}=0} \estunitary^{\reps}(H) U_{\text{p},\xyphase_{ab}=0} |0 \rangle|^2 \label{eq:cosine_circuit} \\
             &= \frac{1}{2}\left(1+\cos\left(\reps \estphase_{ab}\right)\right)~\text{and} \label{eq:cosine_distro} \\
    P_\text{s}(\reps\estphase_{ab}) &= |\langle 0| U^{\dagger}_{\text{p},\xyphase_{ab}=0} \estunitary^{\reps}(H) U_{\text{p},\xyphase_{ab}=\pi/2}|0 \rangle|^2 \label{eq:sine_circuit} \\
             &= \frac{1}{2}\left(1+\sin\left(\reps \estphase_{ab}\right)\right), \label{eq:sine_distro}
\end{align}
\label{eq:distributions}
\end{subequations}
where the circuits that sample from these distributions are evident in \eqnsref{eq:cosine_circuit}{eq:sine_circuit} and the functional forms of the distributions are given in \eqnsref{eq:cosine_distro}{eq:sine_distro}.
Here $\reps$ is the number of applications of $\estunitary(H)$ during the $g$th generation.
As indicated above $\reps$ is chosen with logarithmic spacing, i.e., $\reps = 2^g$, and experiments proceed by refining the estimate of $\estphase_{ab}$ across generations consisting of increasing numbers of repetitions of $\estunitary(H)$~\footnote{Other spacings, i.e., $\reps \neq 2^g$ may potentially be utilized but we do not consider those cases here.}.

For a fixed value of $\reps$, the circuits represented by \eqnsref{eq:cosine_circuit}{eq:sine_circuit} are repeated sufficiently many times to estimate $P_\text{c}$ and $P_\text{s}$ from the relative frequencies of 0 and 1 outcomes.
\eqnsref{eq:cosine_distro}{eq:sine_distro}, then, unambiguously specify $\estphase_{ab}$ on a segment of $2\pi/\reps$ radians,
\begin{equation}
    \reps\estphase_{ab}=\arctantwo\left(2P_\text{c}-1, 2P_\text{s}-1\right)~\text{mod}~2\pi, \label{eq:estphase_to_probs}
\end{equation}
where $\arctantwo$ accounts for the branch cuts of $\arctan$ by tracking the signs of the $x$ and $y$ components.
RPE uses estimates of $\estphase_{ab}$ from experiments with $k_{g'}$ for $g'<g$ to select a particular segment.
At each successive generation, if the right branch is chosen, the error in $\estphase_{ab}$ will exhibit Heisenberg-like scaling.

One of the key features of RPE is its tolerance to additive errors in $P_\text{c}$ and $P_\text{s}$.
In the Supplemental Materials we study the impact of coherent errors on state preparation ($U_{\text{p},\xyphase_{ab}}$) and unpreparation ($U_{\text{p},\xyphase_{ab}}^\dagger$).
The parameters of the error channel under consideration are related to the deviation of the state prepared (or unprepared) relative to the target state, $\ket{\xyphase_{ab}}$.
These include errors that lead to support with erroneous amplitude ($\errc$) and phase ($\errp$) in the ``target subspace'', i.e., $\lsp\lbrace \ket{E_a},\ket{E_b} \rbrace$, but orthogonal to $\ket{\xyphase_{ab}}$.
It also includes leakage errors ($\errl$) that lead to support outside of that subspace.
We indicate the equivalent errors occurring during unpreparation with primed variables (e.g., $\errc'$).

We have derived worst-case bounds on the associated additive contributions to $P_\text{c}$ and $P_\text{s}$, and translated these into worst-case bounds on additive error in the estimate of $\reps \estphase_{ab}$ (see \eqnref{eq:estphase_to_probs}).
This additive error is henceforth indicated as $\delta_\lambda$~\footnote{See Appendix D in the Supplemental Materials.}.
Combined with prior work that establishes the bounds on additive errors under which RPE can succeed~\cite{kimmel2015robust,russo2020unpub_rpe_theory}, we have identified the conditions on coherent SPAM errors that permit estimation of energy differences with Heisenberg-like scaling.
Our results indicate a surprisingly high tolerance to these errors, prominently that there are conditions for which RPE will still succeed even if as much as $\sim13\%$ of the probability in the prepared (unprepared) state leaks outside of the target subspace.

\textit{Results.---}
To verify our RPE protocol for evaluating energy differences in physical simulation we conducted a proof-of-concept experiment through the cloud-based IBM Quantum Experience \cite{cross2018ibm,Qiskit}.
We computed three of the independent pairwise energy differences between the four eigenstates of molecular hydrogen (H$_2$) in a minimal basis along its dissociation curve.
Combined with a knowledge of the trace of the Hamiltonian over this subspace, we are able to reconstruct the energy eigenvalues themselves.
The results of our experiment are illustrated in \figref{fig:h2_shibboleth}, in which it is evident that RPE succeeds in accurately computing these eigenvalues from pairwise differences.
All Hamiltonian (and $k_g$) dependence was precompiled into two- or three-CNOT circuits for this two-qubit demonstration, leading to $k_g$-independent depth circuits of at most $11$ CNOTs.
Details of the circuits that were run are provided in the Supplemental Materials.
We remark that this precompilation approach cannot, of course, be useful at application scale, because the quantum advantage derives comes from performing the Hamiltonain evolution quantum mechanically.
For there to be a ``useful'' demonstration, non-trivial quantum hardware improvements will be required, as is the case for all extant quantum algorithms.

\begin{figure}[t]
\includegraphics[width=\columnwidth]{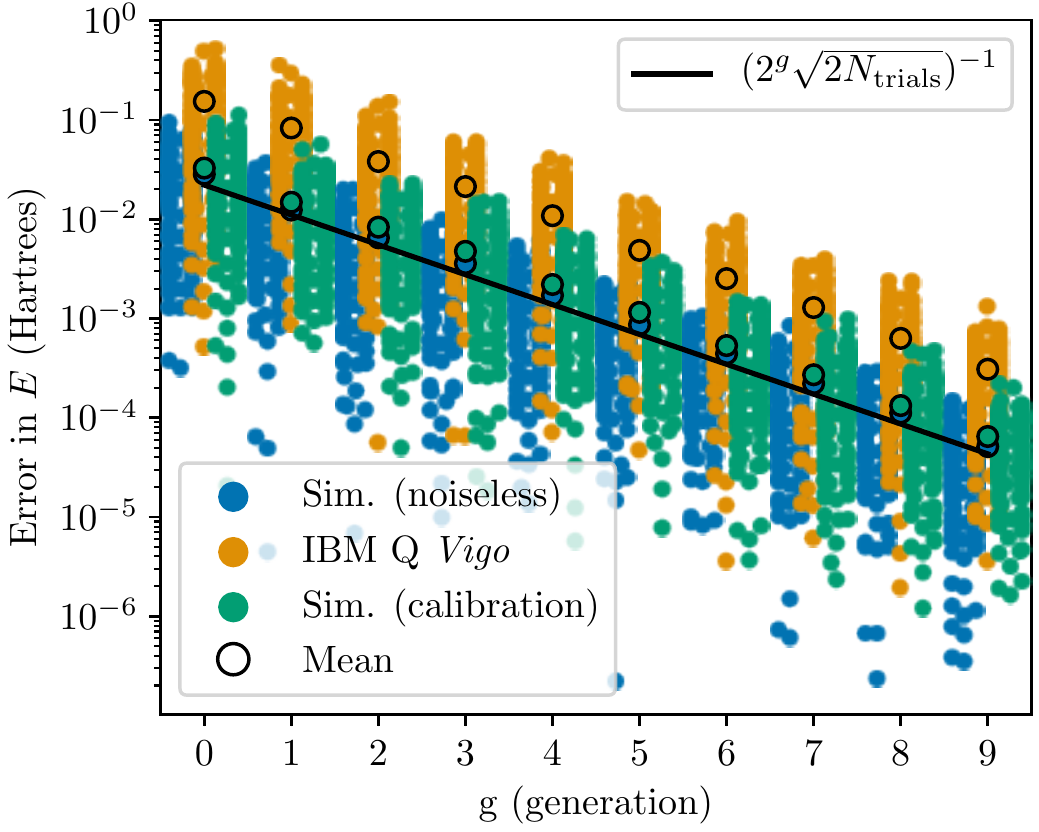}
 \caption{Simulated and experimental distributions of errors in the H$_2$ energy calculation with 1024 repetitions per circuit.
 A swarm plot with errors from all internuclear separations and energy differences, for each generation of RPE.
 Experimental results on IBM Vigo are compared to results from circuit simulations without noise and using the calibration-based noise model supplied by IBM.
 An overall scaling of the error with $1/2^{g}$ is observed, consistent with Heisenberg-like scaling and indicating that the correct branch is predominantly chosen in these sequences.
 \label{fig:h2_shibboleth-error}}
\end{figure}

However, by exploiting the fact that we can compile evolutions for arbitrary $\reps$ into constant-depth circuits we are able to validate that our protocol achieves the ideal scaling with $\reps$.
This is illustrated in \figref{fig:h2_shibboleth-error}, in which we also compare the experimentally observed scaling to the scaling predicted by circuit simulations both with and without noise.
Our noisy circuit simulations are based upon calibration data furnished by IBM at the time of the experiment.
The noiseless simulations provide a benchmark for the optimal performance of our circuits, with the noisy simulations suggesting that experiment will realize a relatively small deviation from this.
The fact that the experiment realizes a mean error that scales with $1/2^{g}$ indicates that we are choosing the correct branch between successive generations, even using noisy hardware.
However, the fact that the noisy simulations predict errors that are almost an order of magnitude smaller than those that are experimentally observed suggests that the furnished noise model is insufficient to predict actual hardware behavior, highlighting both the utility of more expressive noise models \footnote{Such noise models can range from, e.g., one- and two-local completely positive trace-preserving maps \cite{merkel2013self,blume2017demonstration} to ``non-local'' models which include crosstalk errors \cite{rudinger2019probing,sarovar2019detecting} to time-dependent error models \cite{proctor2019detecting}, as opposed to the uniform depolarizing error model implicit in the provided calibration data \cite{magesan2012characterizing,gambetta2012characterization}.}  and the relatively loose relationship between average gate infidelities and worst-case error rates~\cite{sanders2015bounding,kjaergaard2019superconducting}.
Nevertheless, that the procedure still works in the presence of ``hidden'' error processes also highlights RPE's resilience to such ``hidden'' errors.

\begin{figure}[t]
\includegraphics[width=\columnwidth]{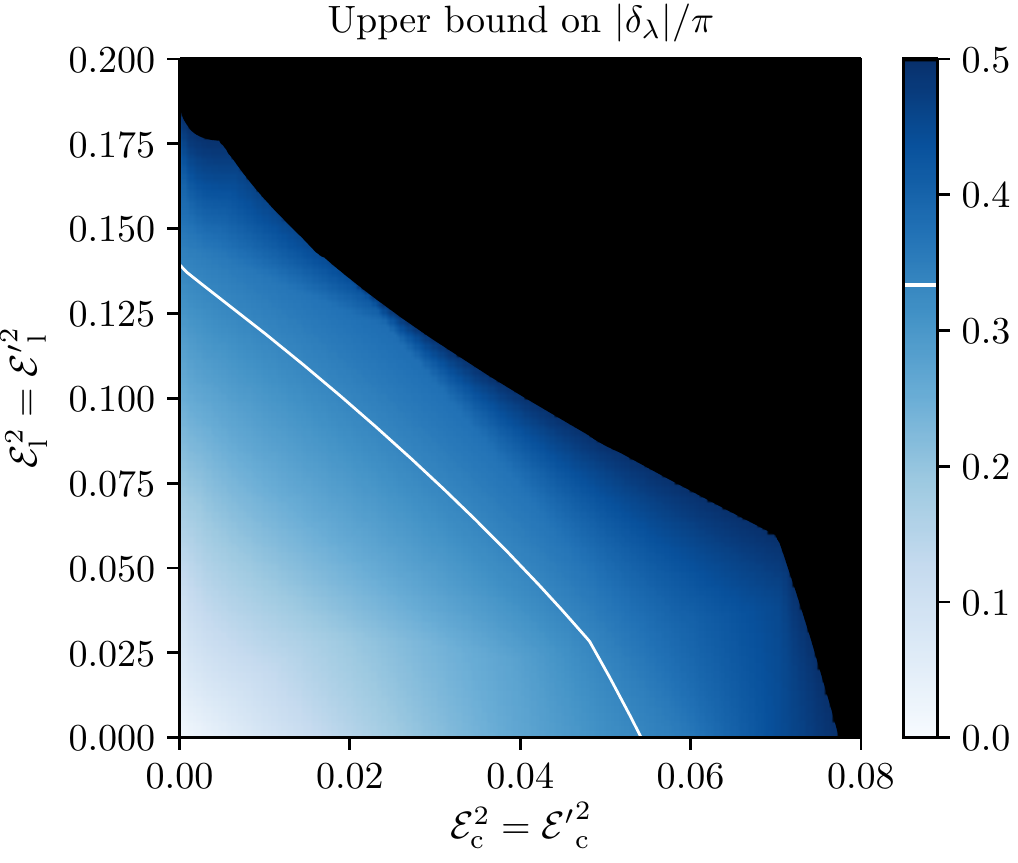}
\caption{Robustness against SPAM errors.
The maximum additive error in the measured angle $\lambda$ used in the RPE protocol, $|\delta_\lambda|$, is plotted as a scaled function of the strength of state preparation errors within ($\errc$) and outside ($\errl$) of the computational subspace.
We have set $\errc=\errc'$ and $=\errl=\errl'$ to get a two-dimensional slice of the error bound in the four-dimensional parameter space.
The upper limit for RPE protocol success, $|\delta_\lambda|<\tfrac{\pi}{3}$, is plotted as a white line \cite{kimmel2015robust,russo2020unpub_rpe_theory}.
The region below this line corresponds to conditions for which RPE will succeed in spite of coherent errors in SPAM.
Values of $|\delta_\lambda|\geq\tfrac{\pi}{2}$ are plotted in black.
The sharp cusps in the cutoff to $\pi/2$ are due to the slack in the error bound, which was optimized for small error values.
Tighter upper bounds would not exhibit this behavior.
\label{fig:robustness}}
\end{figure}

Finally, we illustrate the denominative robustness of RPE to SPAM errors.
\figref{fig:robustness} presents a particular two-dimensional slice of our error model in which $\errc=\errc'$ and $\errl=\errl'$ vary.
All other parameters of the model are optimized over to find a worst-case bound on the additive error in \figref{eq:estphase_to_probs}.
This worst-case additive error is then compared to the upper bound for which the success of RPE is guaranteed.
We find that for $\errc=\errc'=0$, RPE can tolerate a probability of leakage out of the target subspace in each of the preparation and measurement circuits up to $\sim13\%$.
The sensitivity to coherent state preparation errors within the computational subspace is apparently higher, only tolerating individual coherent error probabilities of just $\sim5\%$, partially due to the selection of worst-case phase error within that subspace ($\errp$) in this plot.

\textit{Conclusion.---}
We have demonstrated that RPE can be adapted from its original application in efficiently estimating the phase of a single-qubit gate to efficiently estimating energy differences in quantum simulation.
This approach to phase estimation does not require any auxiliary qubits nor the affiliated controlled implementations of $\estunitary(H)$.
While approaches that use auxiliary qubits can benefit from projection into an energy eigenstate after each round, we have shown that RPE is actually quite tolerant to errors in SPAM.
We expect that the long-term utility of such a protocol is likely to be eclipsed by the auxiliary-qubit-based approaches in future fault-tolerant quantum computers.
However, we do envision this approach as being impactful in the intermediate-term.
Specifically, for verifying and validating quantum simulation algorithms in the era between the noisy, intermediate-scale quantum present and the fault-tolerant quantum error corrected future.

The in-between epoch in which we expect this version of phase estimation to be most useful is one in which the capabilities of quantum computers will be typified by a number of features.
A few error-corrected logical qubits might be available, but with logical error rates and connectivities that are sufficiently limited that the implementation of one-to-many controlled $\estunitary(H)$ is not possible for the desired precision.
There might also be sufficiently few logical qubits that it is possible to classically diagonalize the Hamiltonian over a particular energy window, in which case compilation of the state preparation unitaries from planted solutions will also be possible.
Finally, this protocol might also be useful in diagnosing adiabatic state preparation algorithms which critically rely on finding a pathway between a non-interacting and interacting Hamiltonian in which the ground/first-excited state gap remains as large as possible.
As RPE allows us to efficiently evaluate this gap with limited resources, we see this as one of the more promising applications.

\begin{acknowledgments}
We gratefully acknowledge useful conversations with
Andrew Landahl,
Guang Hao Low, % Kenny discussion regarding arXiv:1904.01131.
Shelby Kimmel, % obviously we've spoken a bunch with Shelby...
Will Kirby, % same w.r.t. Will
Ojas Parekh, % put up with a lot of nonsense from us :)
Nicholas Rubin, % SQuInT
Mohan Sarovar, % also put up with a lot of nonsense from us :)
Rolando Somma, % encouraged us to look at state prep robustness
James Whitfield, % encouraged us to look at state prep robustness
and Nathan Wiebe. % Multiple Kenny discussions
ADB and KMR conceived of this research project.
KMR directed the technical work and ADB directed the writing of the manuscript.
AER and KMR implemented the necessary routines in the Python package \texttt{pyGSTi} \cite{nielsen2019python,nielsen2020probing}, and ran the cloud-based experiments.
BCAM and ADB provided a critical assessment of the derivations.
All authors contributed to the theory, analysis, and writing of this manuscript.
This work was supported in part by the U.S. Department of Energy, Office of Science, Office of Advanced Scientific Computing Research, Quantum Algorithms Team and Quantum Computing Applications Team programs.
Sandia National Laboratories is a multi-mission laboratory managed and operated by National Technology and Engineering Solutions of Sandia, LLC, a wholly owned subsidiary of Honeywell International, Inc., for DOE's National Nuclear Security Administration under contract DE-NA0003525.
This paper describes objective technical results and analysis.
Any subjective views or opinions that might be expressed in the paper do not necessarily represent the views of the U.S. Department of Energy or the United States Government.
\end{acknowledgments}

\bibliography{references}

% Supplemental material starts here
\clearpage
\widetext
\begin{center}
\textbf{\large Supplemental Materials: Evaluating energy differences on a quantum computer with robust phase estimation}
\end{center}

\setcounter{section}{0}
\setcounter{page}{1}
\makeatletter

%%%%%%%%%%%%%%%%%%%%%%%%%%%%%%%%%%%%%%%%%%%%%%%%%%%%%%%%%%%%%%%%%%%%%%%%%%%%%%%%%%%%%%%%%%%%%%%%
\section{Contents}

The Supplemental Materials elaborate on details of some of the central results in the main body of the paper.
\begin{itemize}
    \item Appendix A provides details of the experiment that was conducted on the IBM Quantum Experience, particularly concerning SPAM, compilation, and the circuits that were ultimately executed.
    \item Appendix B describes an error model for SPAM.
    It then derives an explicit expression for $\Delta_\text{c}$, the difference between the ideal and erroneous measurement probabilities for the cosine of the estimated phase, in terms of the error model's parameters.
    \item Appendix C introduces a geometric construction to facilitate analysis and places bounds on the expression for $\Delta_\text{c}$ derived in Appendix B.
    \item Appendix D uses the bounds derived in Appendix C to arrive at a set of constraints under which the success of robust phase estimation (RPE) is guaranteed, as a function of the error model parameters.
    Here we define succeeding as the errors being sufficiently close to zero that Heisenberg-like scaling is still achieved.
    This is used to generate \figref{fig:robustness} in the main body of the paper.
\end{itemize}

%%%%%%%%%%%%%%%%%%%%%%%%%%%%%%%%%%%%%%%%%%%%%%%%%%%%%%%%%%%%%%%%%%%%%%%%%%%%%%%%%%%%%%%%%%%%%%%%
\section{Appendix A: Experimental Details}
\setcounter{equation}{0}
\setcounter{figure}{0}
\renewcommand{\theequation}{A\arabic{equation}}
\renewcommand{\thefigure}{A\arabic{figure}}

In \figsref{fig:h2_shibboleth}{fig:h2_shibboleth-error} we validated the RPE protocol for evaluating energy differences by using IBM's Vigo to compute the full spectrum of a simple exemplary molecule (H$_2$) in a minimal basis set (STO-6G).
The associated Hilbert space dimension is $N=4$ and we used a Bravyi-Kitaev transformation \cite{bravyi2002fermionic,Seeley2012Dec} to create a mapping between fermionic operators on 2 spin orbitals and Pauli operators on 2 qubits.
The Pauli basis Hamiltonian coefficients for the set of internuclear spacings considered were originally extracted from the arXiv version of Ref.~\onlinecite{o2016scalable}.
That we were able to compute the full spectrum and not \emph{just} the energy differences is due to the fact that it was feasible to compute the energy differences for $N-1$ unique pairs of eigenstates and our knowledge of Tr$\left[H\right]$.
Our experiments required us to synthesize circuits for two distinct classes of unitary operators: 1.) $U_{\text{p},\xyphase_{ab}}$ and its adjoint for $a$ and $b$ indexing any given pair $\lbrace \ket{E_a}, \ket{E_b} \rbrace$ and 2.) $\estunitary^{\reps}$ for a given value of $\reps$.
In both cases we exploited the fact that $H$ is classically diagonalizable.

\begin{figure}[b]
\includegraphics{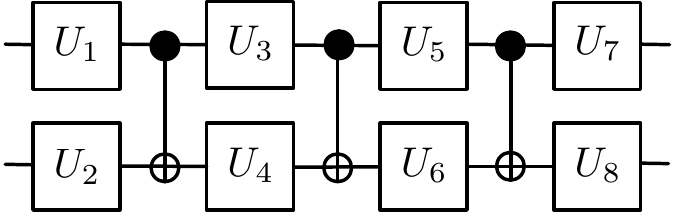}
\caption{
The generic two-qubit circuit template into which $\estunitary^{\reps}$ and $\mathcal{A}(H)$ are compiled.
There exist cases for which it can be done with only two CNOTs.
\label{app_fig:circuit_fig}}
\end{figure}

For any number of qubits $n$ and Hamiltonian $H$, we can synthesize $U_{\text{p},\xyphase_{ab}}$ from a product of two operations:
\begin{equation}
U_{\text{p},\xyphase_{ab}=\beta} = \mathcal{A}(H)\Bselect(a,b,\beta).
\end{equation}
$\mathcal{A}(H)$ maps the computational basis states to energy eigenstates, at least within the target subspace:
\begin{equation}
\mathcal{A}(H)\Big|_{\lsp\{\ket{a},\ket{b}\}} = \ket{E_a}\bra{a} + \ket{E_b}\bra{b}.
\label{eq:O}
\end{equation}
Because the system we studied only requires 2 qubits, we did not try to use adiabatic or filter-based state preparation circuits, or another efficient quantum preparation technique that would be required at application scale.
Instead, we implemented $\mathcal{A}(H)$ as the full unitary transformation of computational basis states to energy eigenstates,
\begin{equation}
\mathcal{A}(H) = \sum_{i} \ket{E_i}\bra{i}.
\label{eq:O_full}
\end{equation}
However, we emphasize that the procedure only relies on the action of $\mathcal{A}(H)$ on $\lsp\{\ket{a},\ket{b}\}$.

$\Bselect(a,b,\beta)$ selects the two computational basis states $\ket{a}$, $\ket{b}$, that correspond to the energy eigenstates of interest $\ket{E_a}$, $\ket{E_b}$, and adds a relative phase $\beta$:
\begin{equation}
\label{eq:Bselect}
\Bselect(a,b,\beta)\ket{0} = \tfrac{1}{\sqrt{2}}\left(\ket{a}+e^{i \beta}\ket{b}\right).
\end{equation}

In general, $\Bselect$ may be implemented with a circuit of depth $O(n)$, and we do so by rotating a control qubit $j$ to have the desired phase (i.e., with a Hadamard followed by a $Z$-rotation by $\beta$), and then entangling that qubit to create the superposition of $\ket{a}$ and $\ket{b}$ states.
See Alg.~\ref{alg:abselect} for a description of \textsc{abSelect}, which performs the entangling step,
\begin{equation}
T\ket{0} = \ket{a}\quad\text{and}\quad T\ket{2^j}=\ket{b}.
\end{equation}
Putting it all together, where $\text{H}$ is the Hadamard,
\begin{equation}
\mathcal{B}(a,b,\beta)=Te^{-i\beta Z_j}\text{H}_j.
\end{equation}

To realize $\estunitary^{\reps}$ for any value of $\reps$ we used classical matrix exponentiation to construct
\begin{equation}
\estunitary^{\reps}(H) = \exp\left(-\frac{i\reps}{\hbar} H\right)
\end{equation}
directly in the computational basis.

To create both the $\estunitary^{\reps}(H)$ and $\mathcal{A}(H)$ circuits, we compiled the necessary two-qubit unitary operations into the native gate set available through the IBM Quantum Experience.
We did so by performing a Quantum Shannon Decomposition in a basis determined by a pair of Cartan decompositions \cite{Drury2008Sep}, as implemented in Qiskit's {\ttfamily two\_qubit\_decompose} module, arriving at two or three CNOT gates and six or eight single-qubit rotations that performed the desired gate \cite{Qiskit}.
The specific single-qubit rotations available are
\begin{equation}
 U_j = \exp\left(-i \alpha_j \hat{n}_j \cdot \vec{\sigma}\right),
\end{equation}
which can be implemented using 3 fixed-axis single-qubit rotations by way of Euler angles.
The form of this circuit is illustrated in \figref{app_fig:circuit_fig}.
Each experiment could have been compiled down even further into 3 CNOTs and 8 single-qubit rotations, but this was not necessary.

\begin{algorithm}[H]
\caption{Choose a control qubit $j$, in-place, and circuit that maps $\ket{0}$ and $\ket{2^j}$ to $\ket{a}$ and $\ket{b}$, respectively.}
\label{alg:abselect}
\begin{algorithmic}[1]
    \item[\textbf{Input:}]
    \\$a,b\in \mathbb{Z}$, the computational basis states to produce
    \item[\textbf{Output:}]
    \\$j\in \mathbb{Z}$, the control qubit $j$ that will select $\ket{a}$ or $\ket{b}$
    \\$T$, the ordered list of quantum instructions execute, i.e., the quantum circuit
    \item[\textbf{Code:}]
\Function{abSelect}{$a$, $b$}
    \State $i\leftarrow 0$ \Comment{The bit currently being processed}
    \State $j\leftarrow -1$ \Comment{The first different bit in $a$ and $b$}
    \State $T\leftarrow []$
    \State $\textrm{flip}\leftarrow 0$
    \State represent $a$ in binary form as $\{ a_k \}_k$
    \State represent $b$ in binary form as $\{ b_k \}_k$
    \While{$i$ is less than the length of $\{ a_k \}_k$ or $\{ b_k \}_k$}
        \If{$a_i=b_i$}
            \If{$a_i=1$}
                \State append $X_i$ to $T$ \Comment{If both classical bits are high, flip qubit $i$ from from $0$ to $1$.}
            \EndIf
        \Else
            \If{$j = -1$}
                \State $j\leftarrow i$ \Comment{Use the first qubit for which the classical bits differ as the control.}
                \State $\textrm{flip} \leftarrow 1-b_i$ \Comment{If the control qubit should be flipped, do so after all CNOTs are processed.}
            \Else
                \State append $\Lambda(X)_{j,i}$ to $T$
                \If{$a_i=1$}
                    \State append $X_i$ to $T$
                \EndIf
            \EndIf
        \EndIf
        \State $n \leftarrow n+1$
    \EndWhile
    \If{$\textrm{flip}$}
        \State append $X_j$ to $T$
    \EndIf
    \State \algorithmicreturn{} $T,j$
\EndFunction
\end{algorithmic}
\end{algorithm}

%%%%%%%%%%%%%%%%%%%%%%%%%%%%%%%%%%%%%%%%%%%%%%%%%%%%%%%%%%%%%%%%%%%%%%%%%%%%%%%%%%%%%%%%%%%%%%%%
\section{Appendix B: Error model for state preparation and measurement}
\setcounter{equation}{0}
\setcounter{figure}{0}
\renewcommand{\theequation}{B\arabic{equation}}
\renewcommand{\thefigure}{B\arabic{figure}}

A feature of approaches to phase estimation with one or more auxiliary qubits is that projective measurement of the auxiliary qubits will leave the system register in an eigenstate of the unitary of which the phase is being estimated.
If there are errors in preparing the precise eigenstate of interest, as long as the prepared state has some overlap with that eigenstate then the probability of projecting into it is proportional to the square of the overlap.
Subsequent rounds of phase estimation can reuse that projected state, mitigating some concerns about the precision of the state preparation unitary.

The RPE protocol described in this paper does not involve any auxiliary qubits.
Instead, each round requires the preparation of a uniform superposition of two eigenstates of the unitary of interest, followed by $\reps$ application of said unitary, and concluded with an ``unprepare'' of the uniform superposition.
(There are, in fact, two distinct circuits per round; one in which the unprepare introduces a relative phase of $\frac{\pi}{2}$ between the supported states, and one in which it does not; these correspond, respectively, to $\beta=\frac{\pi}{2}$ and $\beta=0$, in the application of $U^\dagger_{p,\xyphase_{ab}=\beta}$.)

More precisely, for a unitary $\estunitary$ such that
\begin{subequations}
\begin{align}
\estunitary|E_a\rangle =& e^{i\estphase_a}|E_a\rangle~\forall a \in \lbrace 0,\ldots, N-1 \rbrace~\text{and} \\
\ket{\xyphase_{ab}=\beta} =& \frac{1}{\sqrt{2}} \left(\ket{E_a} +e^{i\beta}\ket{E_b}\right),
\end{align}
\end{subequations}
RPE estimates the difference $\estphase_{ab}=(\estphase_b-\estphase_a)$ mod $2\pi$ using measurements of the probabilities
\begin{equation}
P_c=\left|\bra{\xyphase_{ab}=0}\estunitary^{\reps}\ket{\xyphase_{ab}=0}\right|^2\quad\text{and}\quad
P_s=\left|\bra{\xyphase_{ab}=\tfrac{\pi}{2}}\estunitary^{\reps}\ket{\xyphase_{ab}=0}\right|^2.
\label{app_eqn:probs}
\end{equation}
Because there are no auxiliary qubits, in addition to $\estunitary^{\reps}$ the measurement of these probabilities requires the application of a unitary, $U_{\text{p},\xyphase_{ab}}\ket{0}_{s}=\ket{\xyphase_{ab}}$, and $U_{\text{p},\xyphase_{ab}}^{\dagger}$, for certain values of $\xyphase_{ab}$, for each sample.
It is then natural to be concerned about the impact of errors in $U_{\text{p},\xyphase_{ab}}$ on the estimated phase, $\estphase_{ab}$, independent of errors in implementing $\estunitary$.
In what follows, we provide a model for coherent errors in $U_{\text{p},\xyphase_{ab}}$ and calculate the \emph{exact} error in \eqnref{app_eqn:probs} assuming that model.

There are three real-valued parameters in our model that correspond to distinct coherent errors in $U_{\text{p},\xyphase_{ab}}$:
\begin{enumerate}
    \item $\errc$, the erroneous amplitude on the vector within $\lsp\lbrace \ket{E_a},\ket{E_b} \rbrace$ that is orthogonal to $\ket{\xyphase_{ab}}$.
    \item $\errp$, the erroneous phase on the vector within $\lsp\lbrace \ket{E_a},\ket{E_b} \rbrace$ that is orthogonal to $\ket{\xyphase_{ab}}$.
    \item $\errl$, the erroneous amplitude on the subspace orthogonal to $\lsp\lbrace \ket{E_a},\ket{E_b} \rbrace$, i.e., leakage.
\end{enumerate}
We indicate the non-ideal implementation of $U_{\text{p},\xyphase_{ab}}$ as $\Tilde{U}_{\text{p},\xyphase_{ab}}$ such that,
\begin{equation}
    \Tilde{U}_{\text{p},\xyphase_{ab}=\beta}\ket{0} = \ket{\xyphase_{ab}=\beta;\errc,\errp,\errl}=\frac{\sqrt{1-\errc^2-\errl^2}}{\sqrt{2}} \left( \ket{E_a} + e^{i\beta} \ket{E_b}\right) + \frac{\errc e^{i\errp}}{\sqrt{2}} \left( \ket{E_a} - e^{i\beta} \ket{E_b}\right) + \errl \ket{\errl}.
\end{equation}
We also stipulate that $\braket{\errl|\errl} = 1$ and $\braket{\errl|E_a}=\braket{\errl|E_b}=0$.

To evaluate the \emph{exact} error in \eqnref{app_eqn:probs} subject to this model, it is useful to clarify two notations.
First, \emph{any} quantity with a tilde on top, e.g., $\Tilde{U}_{\text{p},\xyphase_{ab}=\beta}$, corresponds to its non-ideal implementation within our error model.
Second, we will assume that the implementation of $\Tilde{U}^{\dagger}_{\text{p},\xyphase_{ab}=\beta}$ necessary to evaluate the probabilities in \eqnref{app_eqn:probs} might involve different values of the error model parameters.
Accordingly the model parameters for the non-ideal ``prepare'' unitary are indicated without marking and the model parameters for the non-ideal ``unprepare'' unitary are indicated with an apostrophe (e.g., $\errc'$).

The errors that we seek an expression for are
\begin{equation}
\frac{\Delta_{\text{c}}}{2} = \Tilde{P}_{\text{c}}-P_{\text{c}}\quad\text{and}\quad
\frac{\Delta_{\text{s}}}{2} = \Tilde{P}_{\text{s}}-P_{\text{s}},
\label{app_eqn:bare_errors}
\end{equation}
where the factor of 2 accounts for a factor of 1/2 appearing in the probabilities and the subscripts c and s indicate cosine or sine, i.e.,
\begin{equation}
    P_\text{c}=\frac{1}{2}\left[1+\cos(\lambda)\right]\quad\text{and}\quad P_\text{s}=\frac{1}{2}\left[1+\sin(\lambda)\right].
\end{equation}
We only provide a derivation here for $\Delta_\text{c}$ because $\Delta_\text{s}$ can be similarly derived via the identity $\sin(\lambda)=\cos(\pi/2-\lambda)$.
Here $\lambda$ is related to the true phase that we are estimating in any given generation of RPE.
For the $g$th generation, $P_\text{c}$ can be written as
\begin{equation}
    P_\text{c}=|\bra{0} U^{\dagger}_{\text{p},\xyphase_{ab}=0} \estunitary^{\reps} U_{\text{p},\xyphase_{ab}=0}\ket{0}|^2=\frac{1}{2}\left[1+\cos(\reps\theta_{ab})\right],
\end{equation}
such that it is evident that $\lambda=\reps\estphase_{ab}$ and our expression for $\Delta_\text{c}$ will henceforth only implicitly depend on $\reps$ and $\estphase_{ab}$.

$\Tilde{P}_\text{c}$ is given as
\begin{equation}
\Tilde{P}_\text{c}=|\bra{0} \Tilde{U}^{\dagger}_{\text{p},\xyphase_{ab}=0} \estunitary^{\reps} \tilde{U}_{\text{p},\xyphase_{ab}=0}\ket{0}|^2.
\label{app_eqn:P_c_prob}
\end{equation}
We first notice that
\begin{equation}
\estunitary^{\reps}\Tilde{U}_{\text{p},\xyphase_{ab}=\beta}\ket{0} = \estunitary^{\reps}\ket{\xyphase_{ab}=\beta;\errc,\errp,\errl} = \ket{\xyphase_{ab}=\beta+\reps\estphase_{ab};\errc,\errp,\errl},
\end{equation}
up to a global phase and eliding an explicit indication of the change to the vector $\ket{\errl}$, noting that the value of the real parameter $\errl$ is unaffected.
\eqnref{app_eqn:P_c_prob} can then be written as
\begin{equation}
    \Tilde{P}_\text{c} = |\braket{\xyphase_{ab}=0;\errc',\errp',\errl'|\xyphase_{ab}=\lambda;\errc,\errp,\errl}|^2.
\end{equation}
The overlap can be written as
\begin{equation}
    \braket{\xyphase_{ab}=0;\errc',\errp',\errl'|\xyphase_{ab}=\lambda;\errc,\errp,\errl} = \frac{1}{2}A\left( 1+e^{i\lambda}\right) + \frac{1}{2}B\left(1 - e^{i\lambda}\right) + \errl' \errl \left\langle \errl'\middle| \errl\right\rangle,
\end{equation}
where we have introduced the abbreviations
\begin{equation}
A = C'C+\errc'\errc e^{i(\errp-\errp')}
\quad\text{and}\quad B = C \errc' e^{-i\errp'} + C'\errc e^{i\errp},
\end{equation}
where $C^2=1-\errc^2-\errl^2$.
This allows us to write
\begin{multline}
\Tilde{P}_\text{c}=\left|\braket{\xyphase_{ab}=0;\errc',\errp',\errl'|\xyphase_{ab}=\lambda;\errc,\errp,\errl}\right|^2=
\frac{|A|^2}{2}\left( 1+\cos(\lambda)\right) + \frac{1}{2}\Im(AB)\sin(\lambda)
+\frac{|B|^2}{2}\left(1 - \cos(\lambda)\right) \\
+ \errl \errl'\Re \left\{\left( A(1+e^{i\lambda})+B(1-e^{i\lambda})\right)\left\langle \errl'\middle| \errl\right\rangle \right\}
+ \errl^2\errl'^2 \left|\left\langle \errl'\middle|
\errl\right\rangle\right|^2,\label{app_eqn:P_def}
\end{multline}
where $\Im$ and $\Re$ denote, respectively, real and imaginary parts.
We note again the uncontrolled change in $\ket{\errl}$ under the action of $\estunitary$ and indicate that in deriving bounds we will be concerned with a worst-case (maximal) value of $|\Delta_\text{c}|$---in particular we will consider both maximally pessimistic and optimistic choices of $\ket{\errl}$.
This brings us to the main result of this Appendix,
\begin{multline}
\Delta_\text{c} = |A|^2\left( 1+\cos(\lambda)\right) +|B|^2\left(1 - \cos(\lambda)\right) + \Im(AB)\sin(\lambda) - 1 - \cos(\lambda)\\
+ 2\errl\errl'\Re \left\{
\left( A(1+e^{i\lambda})+B(1-e^{i\lambda})\right)\braket{\errl'|\errl} \right\}
+ 2\errl^2\errl'^2 \left|\braket{\errl'|\errl}\right|^2.
\label{app_eqn:error}
\end{multline}

%%%%%%%%%%%%%%%%%%%%%%%%%%%%%%%%%%%%%%%%%%%%%%%%%%%%%%%%%%%%%%%%%%%%%%%%%%%%%%%%%%%%%%%%%%%%%%%%
\section{Appendix C: A Bound on the Error\label{app:error_bound}}
\setcounter{equation}{0}
\setcounter{figure}{0}
\renewcommand{\theequation}{C\arabic{equation}}
\renewcommand{\thefigure}{C\arabic{figure}}

We are interested in determining upper and lower bounds on $\Delta_\text{c}$ as a function of the error model's amplitude parameters, $\errc$, $\errc'$, $\errl$, and $\errl'$.
Specifically, the upper (lower) bound corresponds to a maximally pessimistic (optimistic) choice of the error model's phase parameters, $\errp$, $\errp'$, $\ket{\errl}$, and $\ket{\errl'}$.
To facilitate this analysis it is useful to reorganize \eqnref{app_eqn:error} according to three distinct contributions to the error:
\begin{enumerate}
    \item Coherent errors within $\lsp\lbrace \ket{E_a},\ket{E_b} \rbrace$ that do not depend on $\lambda$.
    \item Coherent errors within $\lsp\lbrace \ket{E_a},\ket{E_b} \rbrace$ that do depend on $\lambda$.
    \item Leakage errors with terms that are both $\lambda$-dependent and not.
\end{enumerate}
To account for the $\lambda$-dependence of the error it will be useful to concomitantly introduce a geometric interpretation of these terms in \eqnref{app_eqn:error}.

Evidently, $\cos(\lambda)$ and $\sin(\lambda)$ can be interpreted as the $(x,y)$ components of a unit vector, $\hat{n} = \left(\cos(\lambda),\sin(\lambda)\right)$.
We can then interpret the coefficients of these terms in $\Delta_\text{c}$ as arising due to the orientation of $\hat{n}$ relative to a vector whose components depend on $A$ and $B$, $\vec{L}=\left(|A|^2-1)-|B|^2, \Im(AB)\right)$.
Reorganizing \eqnref{app_eqn:error} according to this we find
\begin{multline}
\Delta_\text{c} =
\overbrace{\left( (|A|^2-1) + |B|^2 \right)}^{L_0} + \overbrace{\begin{pmatrix}\cos(\lambda)\\\sin(\lambda)\end{pmatrix}}^{\hat n}\cdot \overbrace{\begin{pmatrix}(|A|^2-1)-|B|^2 \\ \Im(AB)\end{pmatrix}}^{\vec L}\\
+ 2\overbrace{\errl \errl'}^D \overbrace{\left|\braket{\errl'|\errl}\right|}^u\left[ \overbrace{\Re \left\{
\left( A(1+e^{i\lambda})+B(1-e^{i\lambda})\right)\frac{\braket{\errl'|\errl}}{\left|\braket{\errl'|\errl}\right|} \right\}}^{F}
+ \overbrace{\errl \errl'}^D \overbrace{\left|\braket{\errl'|\errl}\right|}^u \right].
\label{app_eqn:Delta_geom_vis}
\end{multline}

We begin by considering the extremal values of the leakage contribution in the second line.
It is the case that $D \geq 0$ because our model absorbs the phase into the states.
It is also the case that $0\leq u\leq 1$.
There are then three possible values of $u$ for which the second line, $2Du(F+Du)$, is either minimized or maximized.
These correspond to the imperfect preparation and unpreparation leakage states being perpendicular ($u=0$) or parallel ($u=1$), or if a particular ratio of $D$ and $F$ lies within $\left[0,1\right]$,
\begin{equation}
\partial_u(2Du(F+Du))=DF+2D^2 u=0 \rightarrow u=-\frac{F}{2D}.
\label{app_eqn:zero_derivative}
\end{equation}
This final condition corresponds to a particular conspiratorial choice of $\ket{\errl}$ and $\ket{\errl'}$ that isn't accounted for in the other two more intuitive cases.
We now consider whether these extrema contribute to the maximum or minimum values of $\Delta_\text{c}$.

$\Delta_\text{c}$ is maximized when $u=1$ and the phase of the preparation and unpreparation leakage states, $\braket{\errl'|\errl}$, is chosen to be maximally pessimistic.
This corresponds to the leakage contribution taking on its maximum value, independent of both contributions due to coherent errors within the subspace.
In the worst-case, this means that the phase could conspire to allow $F$ to achieve the magnitude
\begin{equation}
|A(1+e^{i\lambda})+B(1-e^{i\lambda})|
\end{equation}
rather than merely the real part, and with unrestricted sign such that $F>0$.
This gives us the upper bound
\begin{equation}
\Delta_\text{c} \leq L_x + \hat n \cdot \vec L + 2D \left[F_{\text{max}} + D\right]
\end{equation}
where
\begin{multline}
\left|  A(1+e^{i\lambda})+B(1-e^{i\lambda}) \right|
 = \left| (A+B)+(A-B)e^{i\lambda} \right|\leq |A+B|+|A-B| \\
 \leq 2\sqrt{ (C'C+\errc'\errc)^2 + (C \errc' + C'\errc)^2 }
 = F_{\text{max}}.
 \label{app_eqn:Fmax_def}
\end{multline}
The choice to removes the $\lambda$, $\errp$ and $\errp'$-dependence from the leakage terms allows the coherent contribution to errors to be treated separately, significantly simpliflying the analysis, at the expense of making the upper bound slightly looser than it could be.
Note that the $\lambda$-dependence remains in the coherent contribution.

There are two cases to consider in which $\Delta_\text{c}$ might be minimized.
These correspond to whether $F_{\text{max}}\leq 2D$.
Both cases minimize $\Delta_\text{c}$ by choosing the phase of the leakage states such that $F<0$, allowing the leakage contribution to cancel part of the contributions due to coherent errors within the subspace.
Accordingly, the leakage extremum corresponding to $u=0$ does not itself correspond to a maximum or minimum of $\Delta_\text{c}$ because it nullifies the leakage contribution rather than signifying locally maximized leakage.

When $F_{\text{max}}\leq 2D$, the zero derivative condition in \eqnref{app_eqn:zero_derivative} can be met such that the leakage contribution becomes
\begin{equation}
2Du(F+Du)|_{u=-\frac{F}{2D}} = -2D\frac{F}{2D}\left(F-D\frac{F}{2D}\right) = -F\left(F-\frac{F}{2}\right)=-\frac{F^2}{2}.
\end{equation}
Replacing $F$ with the $\lambda$-independent $F_{\text{max}}$ we find the lower bound
\begin{equation}
\Delta_\text{c} \geq L_0 - \hat n\cdot \vec L - \frac{F_{\text{max}}^2}{2}.
\end{equation}
When $F_{\text{max}} > 2D$, the zero derivative condition cannot hold and the leakage contribution is minimized when $u=1$, recalling that we are now considering $F<0$.
This gives us another lower bound,
\begin{equation}
\Delta_\text{c} \geq L_0 + \hat n \cdot \vec L - 2D \left[F_{\text{max}} + D\right].
\end{equation}

Combining the upper and lower bounds and dropping the shorthand for $D$ we find
\begin{equation}
\Delta_{c,\text{min}} = \overbrace{L_0 - \min\left\{\frac{F_{\text{max}}^2}{2},2\errl\errl'[F_{\text{max}}+\errl\errl']\right\}}^{L_-} - \hat n\cdot \vec L
\leq \Delta_{c} \leq \overbrace{L_0 + 2\errl\errl' \left[F_{\text{max}}+\errl\errl' \right]}^{L_+} + \hat n\cdot \vec L =\Delta_{c,\text{max}},\label{app_eqn:error_bound}
\end{equation}
recalling from \eqnsref{app_eqn:Fmax_def}{app_eqn:Delta_geom_vis}
\begin{equation}L_0= |A|^2-1+|B|^2,\quad \vec L = \begin{pmatrix}(|A|^2-1)-|B|^2 \\ |A||B|\end{pmatrix}, \quad\text{and}\quad F_\text{max}= 2\sqrt{ (C'C+\errc'\errc)^2 + (C \errc' + C'\errc)^2 }.
\end{equation}
It is worth emphasizing that $\Delta_{\text{c}}$ is a \emph{signed} quantity.
That is, RPE can fail if $\Delta_{\text{c}}$ is too small, i.e., too negative.
In the following Appendix, we consider bounds on the errors in $\Delta_{\text{c}}$ and $\Delta_{\text{s}}$, along with the geometric interpretation introduced in this Appendix and construct geometric constraints for RPE's success.

%%%%%%%%%%%%%%%%%%%%%%%%%%%%%%%%%%%%%%%%%%%%%%%%%%%%%%%%%%%%%%%%%%%%%%%%%%%%%%%%%%%%%%%%%%%%%%%%
\section{Appendix D: Constraints On the Success of RPE}
\setcounter{equation}{0}
\setcounter{figure}{0}
\renewcommand{\theequation}{D\arabic{equation}}
\renewcommand{\thefigure}{D\arabic{figure}}

\begin{figure}[ht]
\graphicspath{{figures/}}
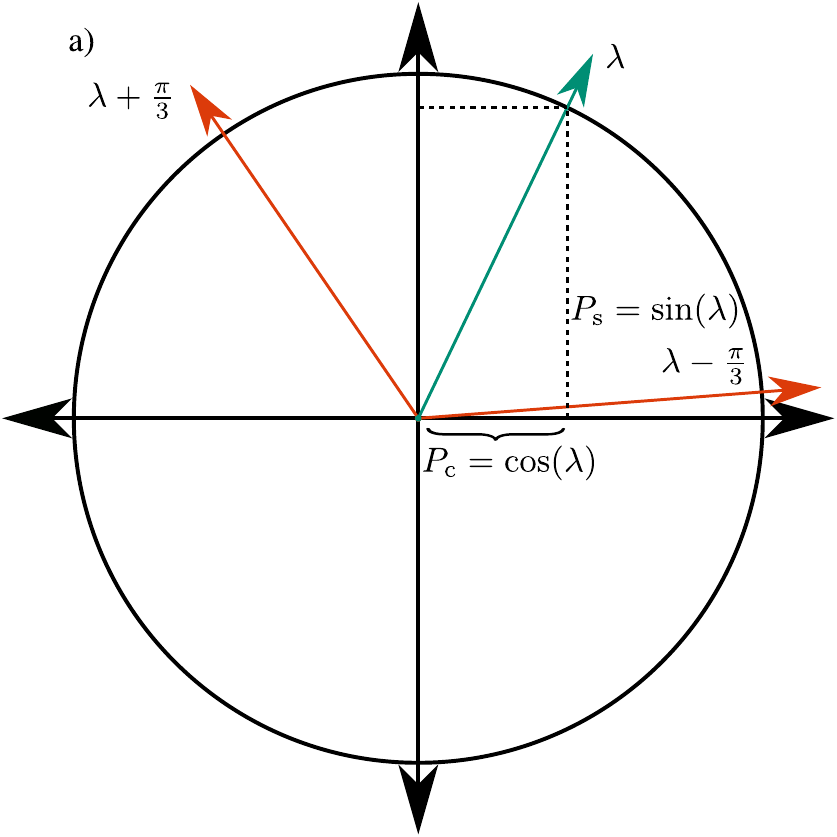
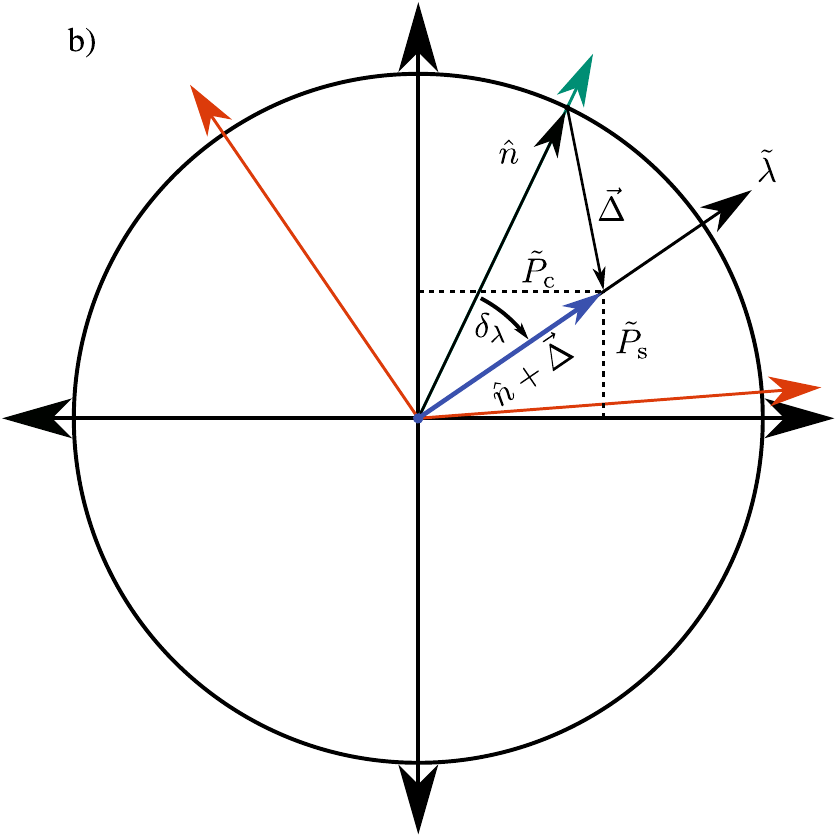
\\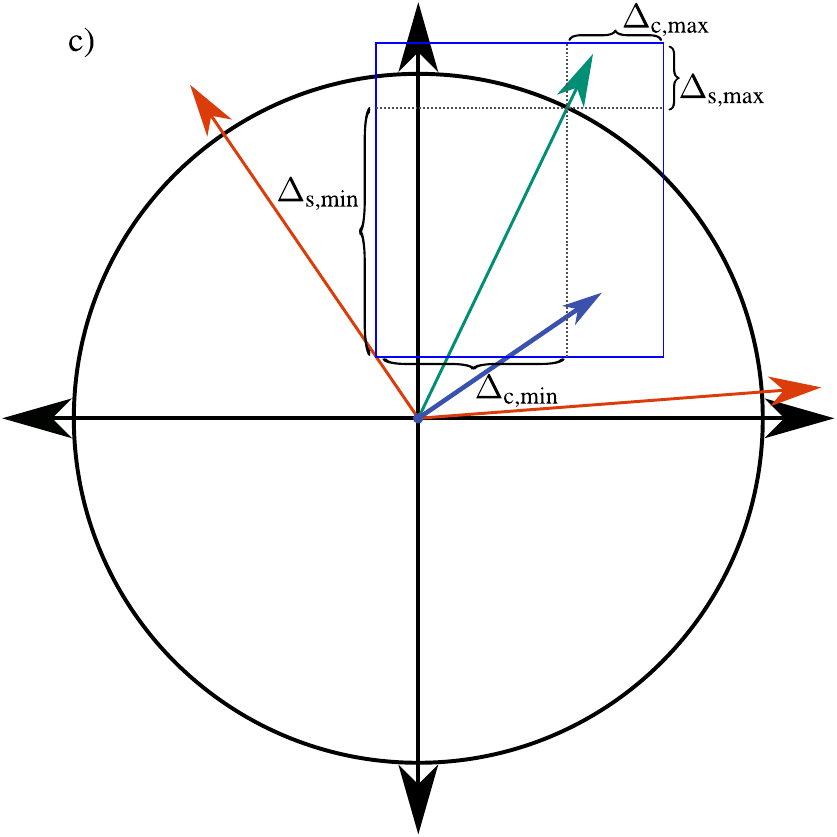
\caption{Bounding the error on the estimated angle, $\tilde{\lambda}$, in the presence of SPAM errors described in Appendix B.
(a) The green arrow represents $\lambda = \arctantwo\left(P_\text{s},P_\text{c}\right) = \arctantwo\left(n_y,n_x\right)$, the angle which would be measured in the absence of errors at generation $g$, $\lambda=k_g\theta_{ab}$.
The red arrows at $\lambda\pm\frac{\pi}{3}$ represent the maximum deviation from this angle that RPE can tolerate while still guaranteeing a correct final result~\cite{russo2020unpub_rpe_theory}.
(b) The presence of SPAM errors induces an additive shift in $\hat n$ by $\vec \Delta$, yielding $(\tilde{P}_c,\tilde{P}_s)=\hat n+\vec \Delta$.
The error in $\tilde{\lambda}$, the angle inferred from this imperfect data, is $\delta_\lambda$.
(c) Appendix C bounded $\vec \Delta$ using the error model's parameters, in turn bounding $\hat n+\vec \Delta$ within the blue box.
In the illustrated case, the bounds saturate the $\tfrac{\pi}{3}$ tolerance of the RPE protocol.
\emph{For this specific} $\lambda$, RPE is therefore guaranteed to succeed.
But, because we do not know what $\lambda$ is, one must consider all possible $\lambda$ to guarantee the success of the RPE protocol.
 \label{app_fig:RPE_fig}}
\end{figure}

The basis for RPE is a nonadaptive phase estimation protocol with Heisenberg-like scaling in which values of $\reps$ are spaced logarithmically and experiments are conducted in order of increasing $\reps$ as to increasingly restrict the range of subsequent estimates of $\estphase_{ab}$~\cite{higgins2009demonstrating}.
This was originally derived in the absence of gate and SPAM errors for a single qubit.
Among the primary advances of RPE is that it achieves Heisenberg-like scaling, even in the presence of sufficiently small additive errors on $P_\text{c}$ and $P_\text{s}$, i.e., if $\max\{|\Delta_\text{c}|,|\Delta_\text{s}|\}/2<1/\sqrt{8}$~\cite{kimmel2015robust}.
The factor of 2 comes from our definition of $\Delta_\text{c}$ and $\Delta_\text{s}$ in \eqnref{app_eqn:bare_errors}.
In fact, we believe that this bound is slightly too optimistic and that the more stringent $\max\{|\Delta_\text{c}|,|\Delta_\text{s}|\}/2<\sqrt{3/32}$ is required~\cite{russo2020unpub_rpe_theory}.
Just prior to making this preprint publicly available we were informed of other authors with similar findings~\cite{belliardo20a}.
We note that the numerical discrepancy is ultimately quite small (i.e., $1/\sqrt{8}\approx 0.354$ and $\sqrt{3/32}\approx 0.306$).
Our analysis proceeds assuming our more restrictive bound.
Here we consider RPE as succeeding if the errors are sufficiently small that Heisenberg-like scaling can still be achieved.

We first define the error vector, $\vec{\Delta} = \left(\Delta_\text{c},\Delta_\text{s}\right)$, noting that it is easy to derive an expression for $\Delta_\text{s}$ from \eqnref{app_eqn:error} using the identity $\sin(\lambda)=\cos(\pi/2-\lambda)$.
Within our error model, RPE will estimate
\begin{equation}
    \tilde{\lambda} = \arctantwo\left(\sin(\lambda)+\Delta_\text{s},\cos(\lambda)+\Delta_\text{c} \right) = \lambda + \delta_{\lambda} = \reps \estphase_{ab} + \delta_{\lambda},
\end{equation}
where $\delta_\lambda$ denotes the discrepancy between $\tilde{\lambda}$ and $\lambda$, i.e., $\delta_\lambda=\tilde{\lambda}-\lambda$.
In terms of our geometric picture, this corresponds to RPE estimating the angle $\hat{n}$ makes with the $x$-axis, using the imperfect $\hat{n} + \vec{\Delta}$.

One could then either follow the procedure from Ref.~\onlinecite{kimmel2015robust} to find the conditions on our error model parameters such that $|\Delta_\text{c}|$ and $|\Delta_\text{s}|$ are both less than $1/\sqrt{2}$, guaranteeing Heisenberg-like scaling, or develop bounds on $|\delta_\lambda|$ that guarantee that these criteria are met.
The first approach follows naturally.
It is straightforward to bound
\begin{equation}
 \hat n\cdot \vec L \leq \sqrt{\left(|A|^2-1-|B|^2\right)^2 +|A|^2|B|^2},
\end{equation}
and subsequently check if both $\Delta_\text{c}$ and $\Delta_\text{s}$ are bounded by $1/\sqrt{2}$.
For the latter procedure, and using our more restrictive bound, it turns out that the equivalent bound is $|\delta_\lambda| < \frac{\pi}{3}$~\cite{russo2020unpub_rpe_theory}.
This more tightly constrains the performance of RPE because it considers the aggregate impact of both $\Delta_\text{c}$ and $\Delta_\text{s}$ on the estimate of interest, rather than bounding each independently.
We make use of our geometric picture to illustrate this in \figref{app_fig:RPE_fig}.
Rather than checking both $|\Delta_\text{c}|$ and $|\Delta_\text{s}|$, we now only need to check $\delta_{\lambda}$, specifically that
\begin{equation}
\frac{\hat n +\vec \Delta}{\left|\hat n +\vec \Delta\right|}\cdot \hat n = \cos(\delta_\lambda) > \cos\left(\frac{\pi}{3}\right)= \frac{1}{2}.
\end{equation}
Noting that the we are now verifying that a quantity is greater than $1/2$ because cosine has even symmetry on $\left(-\pi/3,\pi/3\right)$ and is monotonically decreasing on $\left[0,\pi/3\right)$.

\eqnref{app_eqn:error_bound} provides upper and lower bounds on $\Delta_\text{c}$,
\begin{equation}
\Delta_\text{c} \lessgtr L_\pm \pm \hat n \cdot \vec L.
\label{app_eqn:delta_c_bound}
\end{equation}
Rotating $\hat n$ clockwise by $\pi/2$ gives us $\hat n^\bot=\begin{pmatrix}\sin(\lambda) \\ -\cos(\lambda) \end{pmatrix}$,
which naturally expresses the analogous bound for $\Delta_\text{s}$,
\begin{equation}
\Delta_\text{s} \lessgtr L_\pm \pm \hat n^\bot \cdot \vec L.
\label{app_eqn:delta_s_bound}
\end{equation}
The maximal value of $\delta_\lambda$ will be achieved for $\vec{\Delta}$ saturating these bounds.
These extrema coincide with the vertices of the blue box in \figref{app_fig:RPE_fig}.
The choice of $+$ or $-$ branch in \eqref{app_eqn:delta_c_bound} determines if you are on the right or left side of the box in \figref{app_fig:RPE_fig}.
Similarly, \eqref{app_eqn:delta_s_bound} determines if you are on the top or bottom of the box.
Therefore, for the vertex where both $x$ and $y$ are maximum (minimum), i.e., top-right (bottom-left),
\begin{equation}
\vec \Delta = \overbrace{\begin{pmatrix}L_\pm\\L_\pm \end{pmatrix}}^{\vec \ell_0}\pm \begin{pmatrix}L_x & L_y \\ -L_y & L_x\end{pmatrix}
\begin{pmatrix}\cos(\lambda) \\ \sin(\lambda) \end{pmatrix}
= \vec \ell_0 \pm L_x \hat n \pm L_y\hat n ^\bot
= (\ell_0 \cos(\lambda-\varphi_0) \pm L_x )\hat n + (\ell_0 \sin(\lambda-\varphi_0) \pm L_y)\hat n ^\bot,
\label{app_eqn:Delta_same}
\end{equation}
where $\varphi_0$ is the angle that $\vec \ell_0$ makes with the $x$ axis.
Explicitly,
\begin{equation}
\vec \Delta_\text{top-right} = \overbrace{\begin{pmatrix}L_+\\L_+ \end{pmatrix}}^{\vec \ell_0}+ \begin{pmatrix}L_x & L_y \\ -L_y & L_x\end{pmatrix}
\begin{pmatrix}\cos(\lambda) \\ \sin(\lambda) \end{pmatrix}
= \vec \ell_0 + L_x \hat n + L_y\hat n ^\bot
= (\ell_0 \cos(\lambda-\varphi_0) + L_x )\hat n + (\ell_0 \sin(\lambda-\varphi_0) + L_y)\hat n ^\bot,
\end{equation}
and
\begin{equation}
\vec \Delta_\text{bottom-left} = \overbrace{\begin{pmatrix}L_-\\L_- \end{pmatrix}}^{\vec \ell_0}- \begin{pmatrix}L_x & L_y \\ -L_y & L_x\end{pmatrix}
\begin{pmatrix}\cos(\lambda) \\ \sin(\lambda) \end{pmatrix}
= \vec \ell_0 - L_x \hat n - L_y\hat n ^\bot
= (\ell_0 \cos(\lambda-\varphi_0) - L_x )\hat n + (\ell_0 \sin(\lambda-\varphi_0) - L_y)\hat n ^\bot.
\end{equation}
Similarly, when the $x$ ($y$) component is maximum and the $y$ ($x$) component is minimum, i.e., bottom-right (top-left),
\begin{multline}
\vec \Delta = \overbrace{\begin{pmatrix}L_\pm\\L_\mp \end{pmatrix}}^{\vec \ell_0}\pm \begin{pmatrix}L_x & L_y \\ L_y & -L_x\end{pmatrix}
\begin{pmatrix}\cos(\lambda) \\ \sin(\lambda) \end{pmatrix}
= \vec \ell_0\pm \begin{pmatrix}L_x & -L_y \\ L_y & L_x\end{pmatrix}
\overbrace{\begin{pmatrix}\cos(-\lambda) \\ \sin(-\lambda) \end{pmatrix}}^{\hat n'}
= \vec \ell_0 \pm L_x \hat n' \mp L_y (\hat n') ^\bot\\
= \left[\ell_0 \cos(\lambda-\varphi_0) \pm (\cos(2\lambda)L_x -\sin(2\lambda)L_y) \right]\hat n
 + \left[\ell_0 \sin(\lambda-\varphi_0) \pm (\sin(2\lambda)L_x-\cos(2\lambda)L_y)\right] \hat n ^\bot.
\end{multline}
Making the arguments of the trigonometric functions uniform,
\begin{multline}
\vec \Delta =
(\ell_0 (\cos(\varphi_0)\cos(\lambda)+\sin(\varphi_0)\sin(\lambda)) \pm (L_x (\cos^2(\lambda)-\sin^2(\lambda)) -2L_y \sin(\lambda)\cos(\lambda) ) )\hat n \\
+ (\ell_0 (-\sin(\varphi_0)\cos(\lambda) +\cos(\varphi_0)\sin(\lambda)) \pm (2L_x \sin(\lambda)\cos(\lambda)-L_y(\cos^2(\lambda)-\sin^2(\lambda))))\hat n ^\bot.
\label{app_eqn:Delta_opp}
\end{multline}
Again, explicitly,
\begin{multline}
\vec \Delta_\text{bottom-right} =
(\ell_0 (\cos(\varphi_0)\cos(\lambda)+\sin(\varphi_0)\sin(\lambda)) + (L_x (\cos^2(\lambda)-\sin^2(\lambda)) -2L_y \sin(\lambda)\cos(\lambda) ) )\hat n \\
+ (\ell_0 (-\sin(\varphi_0)\cos(\lambda) +\cos(\varphi_0)\sin(\lambda)) + (2L_x \sin(\lambda)\cos(\lambda)-L_y(\cos^2(\lambda)-\sin^2(\lambda))))\hat n ^\bot,
\end{multline}
and
\begin{multline}
\vec \Delta_\text{top-left} =
(\ell_0 (\cos(\varphi_0)\cos(\lambda)+\sin(\varphi_0)\sin(\lambda)) - (L_x (\cos^2(\lambda)-\sin^2(\lambda)) -2L_y \sin(\lambda)\cos(\lambda) ) )\hat n \\
+ (\ell_0 (-\sin(\varphi_0)\cos(\lambda) +\cos(\varphi_0)\sin(\lambda)) - (2L_x \sin(\lambda)\cos(\lambda)-L_y(\cos^2(\lambda)-\sin^2(\lambda))))\hat n ^\bot.
\end{multline}
Notice that each of the four vertices takes the form $\hat n +\vec \Delta = \Delta_1(\lambda)\hat n+\Delta_2(\lambda)\hat n^\bot$.
For RPE to succeed we require that
\begin{alignat}{3}
&&\frac{1}{2} &< \cos(\delta_\lambda)\nonumber\\
&\iff &\frac{1}{4} &< \cos^2(\delta_\lambda)&\quad\text{and}\quad &|\delta_\lambda| <\pi\nonumber\\
&\iff &\Delta_1^2+\Delta_2^2 &< 4\Delta_1^2&\quad\text{and}\quad &\Delta_1 >0\nonumber\\
&\iff&|\Delta_2| &< \sqrt{3}\Delta_1.
\label{app_eqn:inequality}
\end{alignat}
Notice that $\vec L$ and $L_0$, within $L_\pm$, still have dependence on $A$ and $B$:
\begin{equation}L_0= |A|^2-1+|B|^2,\quad \vec L = \begin{pmatrix}(|A|^2-1)-|B|^2 \\ |A||B|\end{pmatrix},
\end{equation}
where
\begin{equation}
A = C'C+\errc'\errc e^{i(\errp-\errp')}
\quad\text{and}\quad B = C \errc' e^{-i\errp'} + C'\errc e^{i\errp}.
\end{equation}

The inequality in \eqref{app_eqn:inequality} fails to hold when $|\Delta_2|-\sqrt{3}\Delta_1$ exceeds $0$ as a function of the error model parameters.
We now provide worst-case bounds on when that occurs in terms of the phase error parameters ($\errp$ and $\errp'$) and $\lambda$.
Notice that \eqnref{app_eqn:Fmax_def} has already removed the $\errp$ and $\errp'$ dependence from $F_\text{max}$, so all of the remaining phase error dependence is in $L_0$, $L_x$ and $L_y$.
Moreover, these terms enter linearly in \eqnsref{app_eqn:Delta_same}{app_eqn:Delta_opp}.
Therefore, the bounds achieve their worst case on the extrema of $L_0$, $L_x$, and $L_y$, which is precisely when $A=C'C \pm \errc'\errc$ and $B=C\errc'\pm C'\errc$ (or its additive inverse).
We emphasize here that the extremization over $A$ and $B$ must unfortunately be done individually for each of $L_0$, $L_x$, and $L_y$, introducing slack into the bound.
We suspect that this bound can be improved.

Finally, to handle the $\lambda$ dependence, we check if $|\Delta_2|<\sqrt{3}\Delta_1$ on the finitely many extrema where
\[0=\frac{d}{d\lambda}(|\Delta_2|-\sqrt{3}\Delta_1).\]
The sign change at $\Delta_2=0$ does not contribute to the extremization, so we are only concerned with
\begin{equation}
0=\frac{d}{d\lambda}(\pm \Delta_2-\sqrt{3}\Delta_1)= p_\pm.
\label{app_eqn:deriv_cond}
\end{equation}
We need to check this condition for both \eqnref{app_eqn:Delta_same} and \eqnref{app_eqn:Delta_opp}.

This is relatively straightforward for \eqnref{app_eqn:Delta_same} which corresponds to
\[0=\ell_0\left(\pm\cos(\lambda-\varphi_0)+\sqrt{3}\sin(\lambda-\varphi_0)\right),\]
or $\tan(\lambda-\varphi_0)=\mp1/\sqrt{3}$, or $|\lambda-\varphi_0|=\tfrac{\pi}{2}-\tfrac{\pi}{3}$.
Geometrically this is the shortest path from $\hat n$ on the circle in \figref{app_fig:RPE_fig} to either of the red bounding rays.
This will make a right angle with the bounding ray.

The situation with \eqnref{app_eqn:Delta_opp} is less straightforward, but can be handled numerically.
$p_\pm$ is a polynomial in $\cos(\lambda)$ and $\sin(\lambda)$, and by substituting $\sin^2(\lambda)=1-\cos^2(\lambda)$, all terms containing powers of $\sin(\lambda)$ greater than $1$ can be transformed, yielding
\begin{equation}
p_\pm = p_1[\cos(\lambda)] +p_2[\cos(\lambda)]\sin(\lambda),\label{app_eqn:polynomial_def}
\end{equation}
where $p_1$ and $p_2$ are polynomials.
So, for $p_\pm=0$,
\[p_1[\cos(\lambda)]^2=p_2[\cos(\lambda)]^2\sin^2(\lambda) = p_2[\cos(\lambda)]^2(1-\cos^2(\lambda)) \]
or
\begin{equation}
0=p_1[\cos(\lambda)]^2-p_2[\cos(\lambda)]^2(1-\cos^2(\lambda)).
\end{equation}
This polynomial in $\cos(\lambda)$ can be numerically solved with a root finder.
The remaining parameters are then $\errc$, $\errl$, and their primed counterparts.
A slice of $|\delta_\lambda|$ through this 4-dimensional space is plotted in \figref{fig:robustness}.
\end{document}